\documentclass[namedreferences]{solarphysics}
\usepackage[optionalrh,natbib]{spr-sola-addons} 

\usepackage{graphicx}        
\usepackage{amssymb}        
\usepackage{color}           
\usepackage{url}             

\usepackage[pdfborder={0 0 0 },urlcolor=blue,breaklinks]{hyperref}

\usepackage{solaheader}

\newcommand{\vph}{\ensuremath{v_{\mathrm{ph}}}}
\renewcommand{\vec}[1]{{\mathbfit #1}}

\renewcommand{\theenumi}{\roman{enumi})}

\ifx \doiurl    \undefined \def \doiurl#1{\href{http://dx.doi.org/#1}{\textsf{#1}}}\fi
\ifx \adsurl    \undefined \def \adsurl#1{\href{http://adsabs.harvard.edu/abs/#1}{\textsf{#1}}}\fi
\ifx \arxivurl  \undefined \def \arxivurl#1{\href{http://arxiv.org/abs/#1}{\textsf{#1}}}\fi

\begin{document}
\begin{article}
\begin{opening}
\title{Effects of Field-Aligned Flows on Standing Kink and Sausage Modes Supported by Coronal Loops}

\author{S.-X.~\surname{Chen}\sep
        B.~\surname{Li}\sep
        L.-D.~\surname{Xia}\sep
        Y.-J.~\surname{Chen}\sep
        H.~\surname{Yu}
       }
\runningauthor{S.-X. Chen {\it et al.}}
\runningtitle{Standing Modes in Flowing Coronal Loops}

   \institute{Shandong Provincial Key Laboratory of Optical Astronomy and
            Solar-Terrestrial Environment, School of Space Science and Physics,
            Shandong University at Weihai, 264209 Weihai, China
                     email: \url{bbl@sdu.edu.cn} \\
             }

\begin{abstract}
{Fundamental standing modes and their overtones play an important role in coronal seismology. We examine how a significant field-aligned flow affects standing modes supported by coronal loops, modeled here as cold magnetic slabs. Of particular interest are the period ratios of the fundamental to its $(n-1)^{\rm th}$ overtone ($P_1/nP_n$) for both kink and sausage modes, and the threshold half-width-to-length ratio for sausage modes. For standing kink modes, the flow significantly reduces $P_1/nP_n$ in general, the effect being particularly strong for larger $n$ and when the density contrast $\rho_0/\rho_{\rm e}$ between loops and their surroundings is weak. That said, even when $\rho_0/\rho_{\rm e}$ approaches infinity, this effect is still substantial, reducing the minimal $P_1/nP_n$ by up to 13.7\,\% (24.5\,\%) for $n=2$ ($n=4$) relative to the static case, when the Alfv\'en Mach number $M_A$ reaches $0.8$ where $M_A$ measures the loop flow speed in units of the internal Alfv\'en speed. For standing sausage modes, although not
negligible, the flow effect in reducing $P_1/nP_n$ is not as strong. However, the threshold half-width-to-length ratio is considerably larger in the flowing case than its static counterpart. For $\rho_0/\rho_{\rm e}$ in the range $[9, 1024]$ and $M_A$ in the range $[0, 0.5]$, an exhaustive parameter study yields that this threshold is  well fitted by our Equation~(23) which involves the two parameters in a simple way. This allows one to analytically constrain the combination $(\rho_0/\rho_{\rm e}, M_A)$ for a loop with known width-to-length ratio when a standing sausage oscillation is identified therein. It also allows one to further examine the idea of partial sausage modes, and the flow is found to reduce significantly the spatial extent where partial modes are allowed.}
\end{abstract}
\end{opening}

 \vspace{0.2cm}

\section{Introduction}
Combining the measured parameters of the abundant low-frequency waves and oscillations
    in the solar corona
    with magnetohydrodynamic (MHD) wave theory,
    coronal seismology offers the capability for deducing the parameters
    of the structured corona that prove difficult to be directly found~\citep[see, {\it e.g. } the reviews by][]{2000SoPh..193..139R,2005LRSP....2....3N,2008IAUS..247....3R,2009SSRv..149....1N,
2011SSRv..158..167E,2012RSPTA.370.3193D}.
Both slow and fast waves have been found important for seismological purposes.
Regarding slow waves, the observed instances appear
    both as standing modes~\citep[see][and references therein]{2011SSRv..158..397W}
    and in the form of propagating waves~\citep[for recent reviews, see][]{2006RSPTA.364..461D,2007SoPh..246....3B,2009SSRv..149...65D}.
Likewise, propagating fast waves were identified in eclipse measurements of active-region
    loops~\citep{2001MNRAS.326..428W,2002MNRAS.336..747W},
    and in apparently open structures as indicated by the {\it Transition Region and Corona Explorer}
    (TRACE:~\cite{1999SoPh..187..229H})
    measurements of a post-flare supra-arcade~\citep{2005A&A...430L..65V}
    and the more recent {\it Solar Dynamics Observatory}/{\it Atmospheric Imaging Assembly}
    (SDO/AIA: \cite{2012SoPh..275....3P,2012SoPh..275...17L})
    measurements of
    a funnel of loops~\citep{2011ApJ...736L..13L}.
Moreover, standing kink oscillations in coronal loops, directly imaged by TRACE
    and first reported by~\citet{1999Sci...285..862N} and~\citet{1999ApJ...520..880A},
    seem to abound also in loops measured by the \textit{Hinode}
     experiment~\citep[{\it e.g.} ][]{2008A&A...482L...9O,2008A&A...489L..49E}~\citep[also see][for an overview of the instruments]{2007SoPh..243....3K},
    {\it Solar TErrestrial RElations Observatories}/{\it Sun Earth Connection Coronal and Heliospheric Investigation} (STEREO/SECCHI) \citep[{\it e.g.} ][]{2009ApJ...698..397V}~\citep[for an instrument overview, see][]{2008SSRv..136....5K,2008SSRv..136...67H},
    SDO~\citep[{\it e.g.} ][]{2011ApJ...736..102A,2012A&A...537A..49W}, to name but a few missions.

The multiple periods found in a number of loops experiencing standing kink oscillations,
    interpreted as the fundamental mode and its higher order overtones,
    have found extensive seismological applications~\citep[see, {\it e.g.},][for recent reviews]{
2009SSRv..149....3A,2009SSRv..149..199R}.
The observational basis is that the period ratio $[P_1/nP_n]$ tends to deviate from unity, where
    $P_1$ denotes the period of the fundamental mode, and $P_n$ denotes that of its $(n-1)^{\rm th}$ overtone
    with $n=2, 3, \ldots$
The first evidence for such a behavior was presented by~\citet{2004SoPh..223...77V} using TRACE 171\AA\ imaging data,
    where $P_1/2P_2$ was found to be $0.91$ and $0.82$ in the two cases examined therein,
    which were corroborated by a further study of the same events~\citep[Table 2 in][]{2007A&A...473..959V}.
Examining a new event also imaged by TRACE in its 171\AA\ passband,
    this latter study yielded a value for $P_1/2P_2$ of $0.9$.
Actually, periods of even higher-order overtones were also detected both in TRACE data~\citep{2009A&A...508.1485V}
    (see also~\citeauthor{2007ApJ...664.1210D}~\citeyear{2007ApJ...664.1210D})
    and in the {\it Nobeyama RadioHeliograph} (NoRH) data~\citep{2013SoPh..284..559K}.
In the former study, $P_1/2P_2$ and $P_1/3P_3$ were found to be 0.99 and 0.965, respectively.
In the latter, these read $0.83$ and $0.91$.
Among the mechanisms that may contribute to the departure of $P_1/nP_n$ from unity,
    for thin EUV loops it is concluded that the density
    stratification~\citep[{\it e.g.} ][]{2005ApJ...624L..57A,2006A&A...457..707D,2006A&A...460..893M}
    (see also the review by~\citeauthor{2009SSRv..149....3A}~\citeyear{2009SSRv..149....3A})
    and lateral expansion~\citep{2008A&A...486.1015V,2008ApJ...686..694R} along the loop play
    the most prominent role.
The direct observational consequence is that one may deduce the longitudinal density scale height $[H_\rho]$
    using $1-P_1/2P_2$, as was first advocated by~\citet{2005ApJ...624L..57A},
    and may combine $1-P_1/2P_2$ and $1-P_1/3P_3$ to simultaneously determine
    $H_\rho$ as well as the spatial scale for the loop lateral expansion~\citep{2009A&A...508.1485V}.

Fundamental or global sausage modes together with their first overtones
    were also detected in flaring loops with NoRH~\citep{2003A&A...412L...7N,2005A&A...439..727M}
    and in cool H$\alpha$ loops~\citep{2008MNRAS.388.1899S}.
In the former, $P_1/2P_2$ was found to be $\approx 0.82$ with $P_1 \approx 14-17$~seconds and $P_2 \approx 8-11$~seconds.
In the latter, a value of $\approx 0.84$ was found for $P_1/2P_2$ with $P_1 \approx 587$~seconds
    and $P_2\approx 349$~seconds.
Despite these measurements, making use of $1-P_1/2P_2$ pertinent to sausage modes
    is not as popular as in the case of standing kink modes.
Rather, more attention was paid to the cutoff loop width-to-length ratio, only beyond which
    can {trapped} standing sausage modes be supported.
For instance, capitalizing on this cutoff, \citet{2004ApJ...600..458A} deduced that the sausage oscillations
    measured using radio instruments with observing frequencies ranging from $100$~MHz to $1$~GHz
    prior to the 2000s were likely to be confined in a loop segment instead of
    perturbing the entire loop.
{For loops with width-to-length ratios smaller than the cutoff,
    sausage modes are no longer trapped but become leaky.
However, this change of nature does not mean that these modes are not observationally irrelevant:
    the damping timescale of the leaky modes can be sufficiently
    longer than their periods, making their detection possible in oscillating signals
    of coronal loops with realistic parameters, as shown recently
    by~\citet{2012ApJ...761..134N}.
    }

The effects of a field-aligned flow in coronal loops on the standing modes that they support
    were often neglected.
While this is justifiable when the flow speeds
    are well below the Alfv\'en speed, as was assessed for kink modes supported by
    thin tubes~\citep{2010SoPh..267..377R},
    loop flows are not necessarily always weak.
In fact, speeds in the Alfv\'enic regime ($\approx 10^3$~km~s$^{-1}$) associated with explosive
    events have been reported~\citep[{\it e.g.} ][]{2003SoPh..217..267I,2005A&A...438.1099H}.
Alternatively, as was pointed out by~\citet{2011ApJ...729L..22T}, speeds of similar magnitude
    can be independently inferred from
    the spatial distributions of the phases associated with standing kink modes along loops such as those
    reported by~\citet{2010ApJ...717..458V} using TRACE and SOHO data.
If the effects of such a strong flow are neglected, then the loop magnetic field strength
    would be seriously underestimated with the standard seismological practice, by a factor of three to be precise~\citep{2011ApJ...729L..22T}.

The present study is intended to provide a comprehensive investigation into the flow effects
    on standing modes supported by coronal loops modeled by a zero-$\beta$ slab,
    where $\beta$ is the ratio of the thermal to the magnetic pressure.
Similar studies on cylinder-supported kink modes were carried out
    by~\citet{2010SoPh..267..377R} and more recently by~\citet{2013SoPh..tmp..195E}, both
    adopting the thin-tube limit and assuming weak flow speeds well below
    the Alfv\'en speed.
In the slab case, the most relevant one seems to be
    that by~\citeauthor{2011A&A...526A..75M}~(\citeyear{2011A&A...526A..75M}; hereafter MR11)
    where an exhaustive analytical study on the period ratio $[P_1/2P_2]$
    for static zero-$\beta$ slabs was conducted.
Our previous work~\citep{2013ApJ...767..169L} (hereafter Article I) adopted also a slab geometry,
    and examined in detail
    how the flows, which are allowed to reach the Alfv\'enic range,
    influence $P_1/2P_2$ of both kink and sausage modes for slabs with arbitrary width-to-length ratios
    in both a coronal and a photospheric environment.
The present study extends both MR11 and Article I in three ways:
First, as observations are not restricted to the first overtone but show the existence of even higher ones,
    we will examine how the flows affect overtones with $n$ being up to four.
    By doing this, we will show that measuring simultaneously three periods will
    lead one to determine the density contrast $[\rho_0/\rho_{\rm e}]$ and the internal Alfv\'en Mach number $[M_A]$.
Second, as shown by Article~I, the flow effects may be best brought out by examining, say, whether
    the minimum $P_1/nP_n$ is subject to the lower limits expected for static slabs.
We will extend MR11 to overtones of arbitrary order $[n]$
    by establishing the lower limits of $P_1/nP_n$ for static slabs:
    while they are to be analytically derived for slabs with an Epstein density profile,
    we show that they also hold for slabs with density profiles taking a step-function form.
Third, regarding coronal sausage oscillations, Article~I showed that the most prominent influence a flow has
    is to increase the critical width-to-length ratio $[(w/L)_{\rm cutoff}]$ required for standing sausage modes to be trapped.
We will extend this by conducting a parameter study of $(w/L)_{\rm cutoff}$, and come up with
    an analytical fit that depends only on $\rho_0/\rho_{\rm e}$ and $M_A$.
This simple formula will then be applied to demonstrate how the simple fact that a standing sausage mode exists
    in a coronal loop can constrain the combination of $\rho_0/\rho_{\rm e}$ and $M_A$.
In addition, it allows us to further the study by~\citet{2004ApJ...600..458A} of the partial
    sausage modes.

This article is organized as follows:
Section~\ref{sec_static_slab} examines static slabs, with the purpose of establishing
    the behavior of $P_1/nP_n$ in general,
    and its lower limit in particular.
Then Section~\ref{sec_flow_slabs} examines in detail how introducing a field-aligned flow
    affects standing modes, with particular attention paid to
    the period ratios for kink and sausage modes alike,
    as well as the cutoff width-to-length ratio for sausage modes.
Section~\ref{sec_conc} summarizes the present study.

\section{The Static Case}
\label{sec_static_slab}

Let us start with a static slab of length $L$ aligned with the $z$-axis in a Cartesian geometry.
The background magnetic field $[\vec{B}_0 = B_0 \hat{\vec{z}}]$
  is uniform, whereas the background density $[\rho(x)]$ is structured
   along the $x$-direction, resulting in a non-uniform Alfv\'en speed $[v_{\rm A}(x)=\sqrt{B_0^2/4\pi\rho(x)}]$.
Two forms of $\rho(x)$ are examined: one is the Epstein profile~\citep{1958Landau}  while the other is a step-function form.
The former enables a fully analytic exploration of the period ratios of the standing modes supported
   by static slabs~\citep[{\it e.g.} ][]{1995SoPh..159..399N,2003A&A...409..325C,
   2007A&A...461.1149P,2011A&A...526A..75M}.
Exploiting this analytical tractability, we will derive the expressions for the period ratios $[P_1/nP_n]$ associated with
   overtones of arbitrary order $[n]$, their approximations in a number of physically interesting limits,
   and their lower limits.
The purpose is to show that, while these properties are established for this particular profile,
   they apply also to the case of a step-function profile, which is easier to implement numerically,
   and which is the profile to be used when field-aligned flows are introduced.

\begin{figure}
\centering
\includegraphics[totalheight=4.5cm]{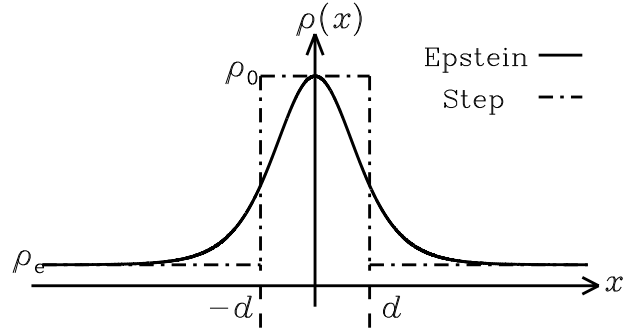}
 \caption{The transverse
density inhomogeneity of a magnetic slab. The solid and dash--dotted curves
correspond to the Epstein and step function profiles, respectively.
}
 \label{fig_slab_illus}
\end{figure}

The Epstein profile is
\begin{equation}\label{epstein}
    \displaystyle\frac{\rho(x)}{\rho_0}=r^{-1}+(1-r^{-1}){\rm sech}^2\displaystyle
    \left(\frac{x}{d}\right) .
\end{equation}
As shown in Figure~\ref{fig_slab_illus}, this density profile continuously connects the asymptotic value $[\rho_{\rm e}]$
   to a maximum $[\rho_0]$ over a characteristic spatial scale $[d]$.
The ratio $[r]$ evaluates the density contrast $[\rho_0/\rho_{\rm e}]$.
Strictly speaking, what Equation~(\ref{epstein}) describes is the symmetric Epstein profile, a particular class of
    the more general Epstein profiles.
It is still referred to as ``the Epstein profile'' throughout only for brevity.
As for the step-function form, it simply reads
\begin{equation}\label{step-rho}
    \rho(x)=\left\{\begin{array}{rcl}
\rho_{\rm e}~,&&~~~~|x|>d \\
\rho_{\rm 0}~.&&~~~~|x|<d
    \end{array}\right.
\end{equation}
In either case, we define $v_{{\rm A}i}$ as $B_0/\sqrt{4\pi\rho_i}$
   with $i=0,~{\rm e}$.
From transverse force balance in zero-$\beta$ MHD it follows that
\begin{equation}\label{density con}
    {\rho_0}/{\rho_{\rm e}} = {v_{\rm Ae}^2}/{v_{\rm A0}^2}.
\end{equation}
Moreover, regardless of the density profile, the half-width of the slab is taken to be $d$,
    resulting in the definition of the aspect ratio as $d/L$.

We restrict ourselves to trapped linear waves that propagate only in the $x$--$z$-plane.
Let $\omega$ and $k$ represent their angular frequency and longitudinal wavenumber.
The phase speed $[\vph]$ is defined as $\vph=\omega/k$.
For standing modes the wavenumber $[k]$ is quantized
\begin{equation}\label{kn}
    k_n=\displaystyle\frac{n\pi}{L}, n=1, 2, \ldots
\end{equation}
This expression is valid for both the fundamental mode $[n=1]$ and its overtones $[n \geq 2]$.
The period ratios $[P_1/nP_n]$ are then simply
\begin{equation}\label{PR-e}
   \displaystyle\frac{P_1}{nP_n} 
   =\displaystyle\frac{k_nv_{{\rm ph},n}}{nk_1 v_{{\rm ph},1}}
   =\displaystyle\frac{v_{{\rm ph},n}}{v_{{\rm ph},1}}~,
\end{equation}
where $v_{{\rm ph},n}\equiv v_{\rm ph}(k_n)$ is the phase speed evaluated at $k_n$.
Hence the $k$-dependence of $\vph$ translates into the dependence on the aspect ratio $[d/L]$ of the period ratios
    $[P_1/nP_n]$.
Not surprisingly, $P_1/nP_n$ depends also on $\rho_0/\rho_{\rm e}$.
Actually, $P_1/nP_n$ is determined by these two parameters, and these two only.

\subsection{The Epstein Profile Slab}
In this case, for kink waves the dispersion relation reads~\citep[{\it e.g.} ][]{1995SoPh..159..399N}
\begin{equation}\label{e-dr-kink}
    \displaystyle\frac{v^2_{\rm ph}}{v^2_{\rm A0}}=1+
    \displaystyle\frac{\displaystyle\sqrt{r^{-2}
    +4k^2d^2-4k^2d^2r^{-1}}-r^{-1}}{2k^2d^2}
    ~,
\end{equation}
   which leads to a general expression for the period ratios
\begin{equation}\label{pr-e-kink}
\left(\displaystyle\frac{P_1}{nP_n}\right)^2=
    \displaystyle\frac{1}{n^2}\left(\displaystyle\frac
    {\displaystyle\frac{2n^2\pi^2d^2}{L^2}-r^{-1}+\sqrt{r^{-2}
    +\displaystyle\frac{4n^2\pi^2d^2}{L^2}
    -\displaystyle\frac{4n^2\pi^2d^2}{L^2}r^{-1}}}
    {\displaystyle\frac{2\pi^2d^2}{L^2}-r^{-1}+\sqrt{r^{-2}
    +\displaystyle\frac{4\pi^2d^2}{L^2}
    -\displaystyle\frac{4\pi^2d^2}{L^2}r^{-1}}}\right)~.
\end{equation}
By deriving Equation~(\ref{pr-e-kink}) we have presented an expression for $P_1/n P_n$
    that is valid for overtones of arbitrary order.
Specializing to the first overtone ($n=2$), one readily recovers Equation~(20) in MR11.
We note, however, the symbol $L$ therein is equivalent to $L/2$ in this study.

Let us proceed by examining the slender- and wide-slab limits.
In the former case $[d/L \ll 1]$, the period ratio
   $P_1/nP_n$ as given by Equation~(\ref{pr-e-kink}) may be satisfactorily
   approximated by
\begin{equation}\label{kink0slim}
    \displaystyle\frac{P_1}{nP_n}\approx1-\displaystyle\frac{\pi^2    d^2\mu_2}
    {2L^2}
    \left(n^2-1\right)-\displaystyle\frac{\pi^4    d^4\mu_2^2}
    {8L^4}
    \left(n^4+2n^2-3\right)~,~~~~~~~~\mu_2\equiv
    \left(\displaystyle\frac{v_{\rm Ae}^2}{v_{\rm A0}^2}-1\right)^2
    ~,
\end{equation}
When the opposite is true $[d/L \gg 1]$, $P_1/n P_n$ may be expressed as
\begin{equation}\label{kink0wide}
    \displaystyle\frac{P_1}{nP_n}\approx1-\displaystyle\frac{L\mu_1}{2\pi
    d}\left(1-\displaystyle\frac{1}{n}\right)+\displaystyle\frac{L^2\mu_1^2}
    {4\pi^2    d^2}\left(\displaystyle\frac{3}{2}-\displaystyle\frac{1}{n}-
    \displaystyle\frac{1}{2n^2}\right)~,~~~~~~~~\mu_1\equiv\displaystyle\sqrt{1-
    \displaystyle\frac{v_{\rm A0}^2}{v_{\rm Ae}^2}} .
\end{equation}
Equations~(\ref{kink0slim}) and (\ref{kink0wide}) generalize Equations~(24) and~(25)
    in MR11 by providing
    the expressions for arbitrary $n$.

The lower limit of period ratios $P_1/nP_n$ turns out to be of
   particular interest, and can be derived in two steps.
First, one notices that $P_1/nP_n$
   at any positive $d/L$ decreases with increasing $r$ as long as $r>1$. Second,
$P_1/nP_n$ in the limit $r\to \infty$ has the asymptotic form
\begin{equation}\label{infty-kink}
    \lim_{r\to\infty}\displaystyle\frac{P_1}{nP_n}
    =\displaystyle\sqrt{\displaystyle\frac{\displaystyle\frac{\pi d}{L}
    +\displaystyle\frac{1}{n}}{\displaystyle\frac{\pi
    d}{L}+1}}>\displaystyle\frac{1}{\sqrt{n}}~
\end{equation}
for any finite aspect ratio $[d/L]$.
Hence the lower limit for $P_1/nP_n$ is $1/\sqrt{n}$, attained when the density contrast approaches infinity
    and the aspect ratio approaches zero.

For sausage waves, the dispersion relation reads~\citep[{\it e.g.} ][]{2007A&A...461.1149P,2009A&A...503..569I}
\begin{equation}\label{e-dr-sausage}
    \displaystyle\frac{v^2_{\rm ph}}{v^2_{\rm A0}}=1+
    \displaystyle\frac{3\displaystyle\sqrt{9r^{-2}-8r^{-1}
    +4k^2d^2-4k^2d^2r^{-1}}-9r^{-1}+4}{2k^2d^2}.
\end{equation}
Consequently, one finds a general expression for the period ratios
\begin{equation}\label{pr-e-sausage}
    \left(\displaystyle\frac{P_1}{nP_n}\right)^2=
    \displaystyle\frac{1}{n^2}\left(\displaystyle\frac
    {\displaystyle\frac{2n^2\pi^2d^2}{L^2}+4-9r^{-1}+3\sqrt{9r^{-2}-8r^{-1}
    +\displaystyle\frac{4n^2\pi^2d^2}{L^2}
    -\displaystyle\frac{4n^2\pi^2d^2}{L^2}r^{-1}}}
    {\displaystyle\frac{2\pi^2d^2}{L^2}+4-9r^{-1}+3\sqrt{9r^{-2}-8r^{-1}
    +\displaystyle\frac{4\pi^2d^2}{L^2}
    -\displaystyle\frac{4\pi^2d^2}{L^2}r^{-1}}}\right) ,
\end{equation}
    which generalizes Equation~(33) in MR11, where only the first overtone is concerned,
    to overtones with arbitrary order $[n]$.

Concerning sausage waves, a number of analytically tractable properties are of interest.
The first concerns the dispersion behavior in the wide-slab limit $[d/L \gg 1]$, in which case
    $P_1/n P_n$ reads
\begin{equation}\label{sausage0wide}
    \displaystyle\frac{P_1}{nP_n}\approx1-\displaystyle\frac{3L\mu_1}{2\pi
    d}\left(1-\displaystyle\frac{1}{n}\right)-\displaystyle\frac{L^2}
    {\pi^2    d^2}\left(1-\displaystyle\frac{1}{n^2}\right)
    +\displaystyle\frac{9L^2\mu_1^2}
    {4\pi^2    d^2}\left(\displaystyle\frac{3}{2}-\displaystyle\frac{1}{n}-
    \displaystyle\frac{1}{2n^2}\right)~,
\end{equation}
where $\mu_1$ is the same as in Equation~(\ref{kink0wide}).
It is noteworthy that Equation~(36) in MR11 is a special case $[n=2]$ of Equation~(\ref{sausage0wide}).
The second concerns the lower bound that $P_1/n P_n$ may attain.
One first notices that $P_1/n P_n$ as a function of $(d/L, r)$ decreases with $r$ at any positive $d/L$
    when $r>1$.
Furthermore, when the density contrast $r\to \infty$, one finds that
\begin{equation}\label{infty-sausage}
    \lim_{r\to\infty}\displaystyle\frac{P_1}{nP_n}
    =\displaystyle\sqrt{\displaystyle\frac{\displaystyle\frac{2\pi^2 d^2}{L^2}
    +\displaystyle\frac{4}{n^2}+\displaystyle\frac{6\pi d}{nL}}
    {\displaystyle\frac{2\pi^2 d^2}{L^2}+4+\displaystyle\frac{6\pi d}{L}}}
    >\displaystyle\frac{1}{n}~.
\end{equation}
Finally, a cutoff wavenumber exists~\citep[{\it e.g.} ][]{2007A&A...461.1149P,2009A&A...503..569I}
\begin{equation}\label{e-cutoff}
    kd=\sqrt{\displaystyle\frac{2}{r-1}}~,
\end{equation}
below which the waves are no longer trapped.
Going a step further, one finds an aspect ratio cutoff
\begin{equation}\label{aspect-cutoff-e}
    (d/L)_{\rm cutoff} = \displaystyle\frac{1}
    {\pi}\sqrt{\displaystyle\frac{2}{r-1}} .
\end{equation}

\subsection{The Step Function Slab}
In this case the dispersion relation reads~\citep[{\it e.g.} ][]{1982SoPh...76..239E,1988A&A...192..343E}
\begin{equation}\label{step-kink}
    \cot\left(kd\sqrt{\displaystyle\frac{v^2_{\rm ph}-v^2_{\rm A0}}{v^2_{\rm
    A0}}}\right)=\displaystyle\frac{v_{\rm Ae}}{v_{\rm
    A0}}\sqrt{\displaystyle\frac{v^2_{\rm ph}-v^2_{\rm A0}}{v^2_{\rm
    Ae}-v^2_{\rm ph}}}
\end{equation}
for kink waves, and
\begin{equation}\label{step-sau}
    \tan\left(kd\sqrt{\displaystyle\frac{v^2_{\rm ph}-v^2_{\rm A0}}{v^2_{\rm
    A0}}}\right)=-\displaystyle\frac{v_{\rm Ae}}{v_{\rm
    A0}}\sqrt{\displaystyle\frac{v^2_{\rm ph}-v^2_{\rm A0}}{v^2_{\rm
    Ae}-v^2_{\rm ph}}}
\end{equation}
for sausage waves.
As in the case where the Epstein profile is adopted, for sausage waves to be trapped,
    the longitudinal wavenumber has to be larger than a cutoff~\citep[{\it e.g.} ][]{1988A&A...192..343E}
\begin{equation}\label{step-cutoff}
    kd=\displaystyle\frac{\pi}{2}\sqrt{\displaystyle\frac{1}{r-1}} .
\end{equation}
Consequently, a cutoff aspect ratio exists
\begin{equation}\label{aspect-cutoff-s}
    (d/L)_{\rm cutoff} = \displaystyle\frac{1}{2}
  \sqrt{\displaystyle\frac{1}{r-1}}~,
\end{equation}
    only beyond which can standing sausage modes be trapped.

Evaluating the period ratios $[P_1/nP_n]$ in the step function case can in principle be done in the same way as in
    the previous case, the only difference being that the phase speed $[\vph]$ has to be found numerically.
Having found $\vph$ as a function of the dimensionless longitudinal wavenumber $[kd]$, one readily finds $P_1/n P_n$
    by employing Equation~(\ref{kn}).

\subsection{Comparing Slabs with Two Different Density Profiles}

Now let us evaluate how sensitively the period ratios depend on the profile that the density inhomogeneity adopts.
In addition, let us examine whether the lower limits of the period ratios are different for different choices of the density profiles.

\begin{figure}
\centering
\includegraphics[totalheight=5.5cm]{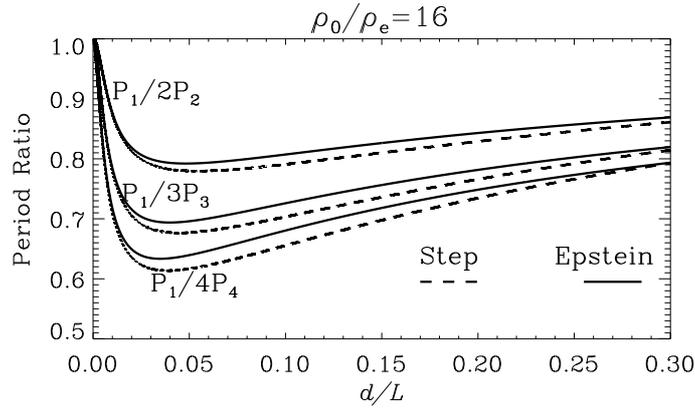}
 \caption{Effects of transverse density profile on standing kink modes supported by static coronal slabs.
Period ratios $[P_1/nP_n (n=2, 3,4)]$ are displayed as functions of aspect ratio
$[d/L]$ for a density contrast $\rho_0/\rho_{\rm e}=16$.
The solid and dashed lines correspond to the Epstein and step-function profiles, respectively.}
 \label{fig_kink_pr_givenR}
\end{figure}

Figure~\ref{fig_kink_pr_givenR} compares the period ratios $[P_1/nP_n]$ for standing kink modes for the Epstein
   (solid curves) and
   step-function density profiles (dashed).
Here the density contrast $[\rho_0/\rho_{\rm e}]$ is $16$, arbitrarily chosen but observationally realistic.
Consider the solid curves first, in which case the behavior of $P_1/nP_n$ can be readily understood
   with the aid of the approximate expressions valid for the Epstein profile.
As expected from Equation~(\ref{kink0slim}), when $d/L$ is small,
   $P_1/nP_n$ starts from unity, regardless of the values of $n$ and density contrast $[\rho_0/\rho_{\rm e}]$.
With increasing $d/L$, $P_1/nP_n$ decreases to its minimum,
   and then approaches unity from below as would be expected from Equation~(\ref{kink0wide}).
Comparing the dashed with the solid curves, one sees that
   choosing different density profiles does not make a qualitative difference.
In fact, the curves corresponding to the same order $[n]$ differ little from one another, with $P_1/nP_n$
   being slightly lower in the step function case.

\begin{figure}
\centering
\includegraphics[totalheight=5.0cm]{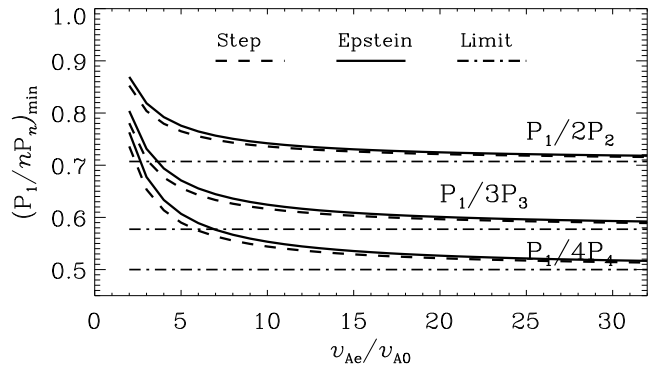}
 \caption{
 Effects of transverse density profile on standing kink modes supported by static coronal slabs.
 Minima of period ratios $[(P_1/nP_n)_{\rm min} (n=2, 3,4)]$ are displayed as functions of the Alfv\'en speed ratio
    $[v_{\rm Ae}/v_{\rm A0}=(\rho_0/\rho_{\rm e})^{1/2}]$.
The solid and dashed lines stand for the Epstein and step-function profiles, respectively.
In addition, the dot--dashed lines represent $1/\sqrt{n}$,
    the lower limits of period ratios $P_1/nP_n$ analytically derived
    for the Epstein profile (Equation~(\ref{infty-kink})).
    }
 \label{fig_static_KS_pr_min}
\end{figure}

Can the slight difference in $P_1/nP_n$ be reflected in their minima $[(P_1/nP_n)_{\rm min}]$ at different
   values of density contrast?
This is examined in Figure~\ref{fig_static_KS_pr_min} where the Alfv\'en speed ratio $[v_{Ae}/v_{A0}]$
   ranges from 2 to 32.
In addition to $(P_1/nP_n)_{\rm min}$ for the two profiles,
   the lower limit $\sqrt{1/n}$ as expected from Equation~(\ref{infty-kink}) is plotted by
   the dash--dotted lines for comparison.
One can see that at all of the examined $v_{Ae}/v_{A0}$, $(P_1/nP_n)_{\rm min}$ for one profile differs little
   from that for the other.
In fact, {a fractional change of at most $3.5\,\%$} is found for all the examined $n$.
Looking further at the period ratios at large $v_{Ae}/v_{A0}$,
   one sees that $(P_1/nP_n)_{\rm min}$ tends to the lower bound [$\sqrt{1/n}$].
Even though it is analytically established for an Epstein profile,
   it holds for the step-function profile as well.

\begin{figure}
\centering
\includegraphics[totalheight=5.5cm]{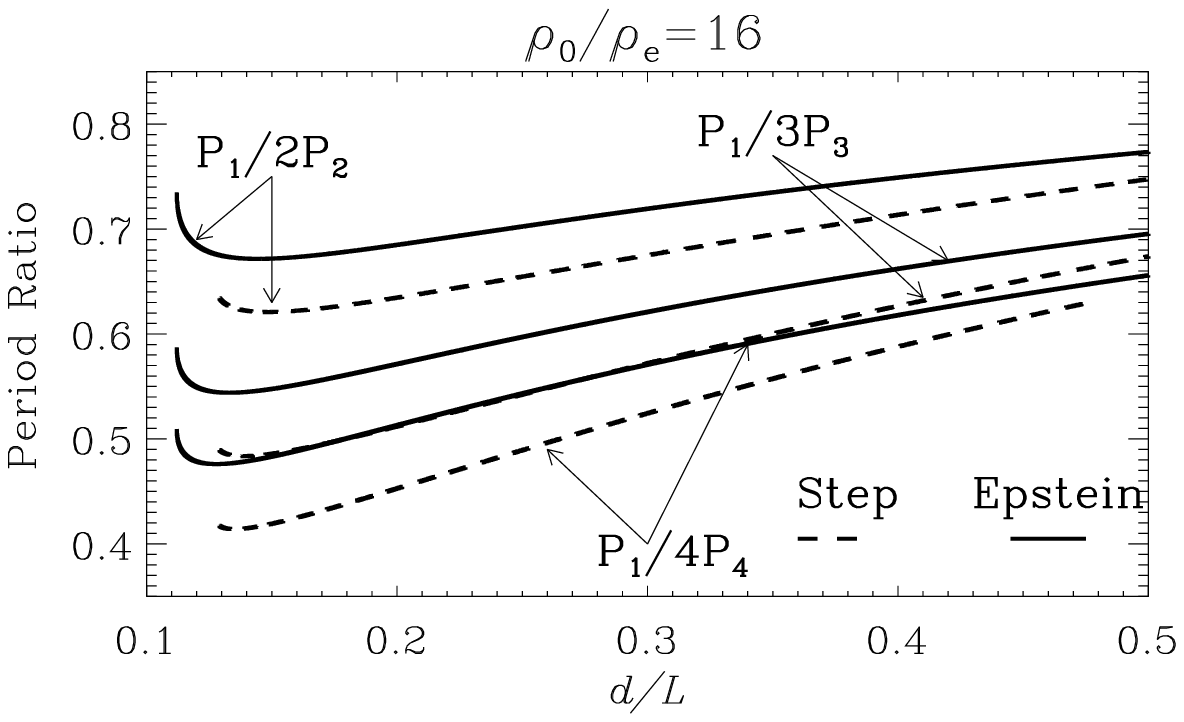}
 \caption{
 Effects of transverse density profile on standing sausage modes supported by static coronal slabs.
Period ratios $[P_1/nP_n (n=2, 3,4)]$ are displayed as functions of aspect ratio
$[d/L]$ for a density contrast $\rho_0/\rho_{\rm e}=16$.
The solid and dashed lines correspond to the Epstein and step function profiles, respectively.
}
 \label{fig_saus_pr_givenR}
\end{figure}

Moving on to the standing sausage modes pertinent to static slabs,
   Figure~\ref{fig_saus_pr_givenR} examines the dependence on the slab aspect ratio $[d/L]$
   of the period ratios $[P_1/nP_n]$ with $n$ being up to four.
It can be seen that the profile-associated difference in $P_1/nP_n$
   is more pronounced than for standing kink modes, with the step-function profile
   corresponding to lower $P_1/nP_n$.
This difference decreases with increasing $d/L$.
The fractional change in $P_1/nP_n$ for the step-function profile relative to the Epstein one
   is found to be {up to $8\,\%$, $10\,\%$, $11\,\%$ for $n=2, 3$, and $4$, respectively.
In this regard, one would have to measure the periods with an accuracy of better
   than $\approx 5\,\%$}
   to tell which profile better describes the density structuring.
One further notes that at this density contrast, $P_1/nP_n$ for both profiles remains substantially larger than
   $1/n$, the expected lower bound as given by Equation~(\ref{infty-sausage}).
On the other hand, one can see that the cutoff aspect ratios $[(d/L)_{\rm cutoff}]$ associated with the two profiles
   are not significantly different, reading {$0.116$ ($0.129$)} for the Epstein (step-function) profile.
Actually, analytically expected values for these, Equations~(\ref{aspect-cutoff-e}) and (\ref{aspect-cutoff-s}),
   yield that the ratio between the two is $2\sqrt{2}/\pi\approx 0.9$.

\begin{figure}
\centering
\includegraphics[totalheight=5.5cm]{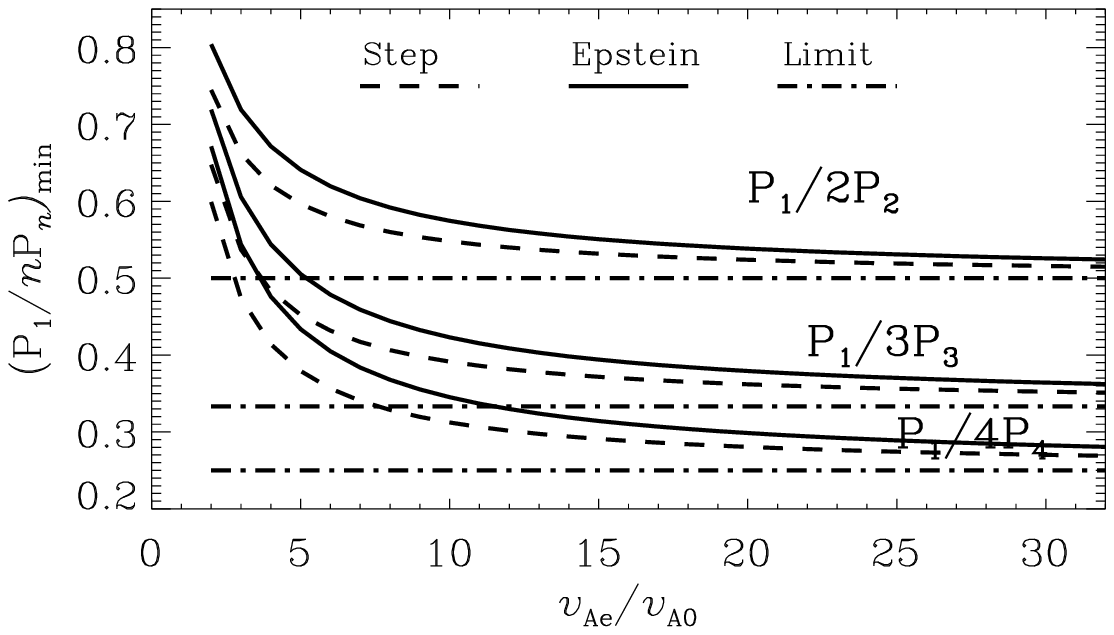}
 \caption{
 Effects of transverse density profile on standing sausage modes supported by static coronal slabs.
 Minima of period ratios $[(P_1/nP_n)_{\rm min} (n=2, 3,4)]$ are displayed as functions of the Alfv\'en speed ratio
    $[v_{\rm Ae}/v_{\rm A0}=(\rho_0/\rho_{\rm e})^{1/2}]$.
 The solid and dashed lines stand for the Epstein and step-function profiles, respectively.
 The dash--dotted lines represent $1/n$, the lower limits of $P_1/nP_n$
     analytically expected for the Epstein profile.}
 \label{fig_static_Saus_pr_min}
\end{figure}

Figure~\ref{fig_static_Saus_pr_min} examines how the minima of $P_1/nP_n$ for standing sausage modes depend on
   the choice of the density profiles for
   Alfv\'en-speed ratios $v_{Ae}/v_{A0}$ ranging from $2$ to $32$.
For comparison, the lower bounds $[1/n]$ expected analytically from Equation~(\ref{infty-sausage})
   are plotted by the dash--dotted lines.
One can see that while the $(P_1/nP_n)_{\rm min}$ curves for the step-function profile (the dashed curves)
   lie generally below those for the Epstein profile (solid),
   the difference between them decreases with increasing $v_{Ae}/v_{A0}$.
The fractional changes between the two, the step function case relative to the Epstein one,
   are {typically $10\,\%$} for all values of $n$ considered when $v_{Ae}/v_{A0} <15$,
   but drop {below $6\,\%$} when $v_{Ae}/v_{A0} >20$.
For both profiles, $(P_1/nP_n)_{\rm min}$ tends to $1/n$ when $v_{Ae}/v_{A0}$ is sufficiently large,
   meaning once again that while this lower bound is established analytically for the Epstein profile,
   it is valid also for the step-function one.

\section{The Flowing Case}
\label{sec_flow_slabs}

In this section, we will examine the effect of field-aligned flow on
   the standing modes
   supported by a magnetic slab.
For the ease of numerical implementation, a step-function form is chosen for the transverse density profile,
   as well as for the speed profile.
The flow speed external to the slab $[U_{\rm e}]$ is taken to be zero, but
   the internal flow $[U_0]$ is in general finite, and
   this is conveniently measured in units of the internal Alfv\'en speed.
In other words, $U_0\equiv M_{\rm A}v_{\rm A0}$ with
    $M_{\rm A}$ being the internal Alfv\'enic Mach number.

The dispersion relation of trapped linear waves supported by a cold slab
   with field-aligned flow can be found by letting the sound speeds approach zero
   in the more general versions~\citep[{\it e.g.} ][]{1995SoPh..159..213N}.
One finds
\begin{equation}\label{flow-dis}
    \displaystyle\frac{\rho_{\rm e}}{\rho_0}
    \displaystyle\frac{n_0}{|m_{\rm e}|}
    \displaystyle\frac{v_{\rm Ae}^2-v^2_{\rm ph}}
    {v_{\rm A0}^2-\left(v_{\rm ph}-U_0\right)^2}=\left\{\begin{array}{c}
                                                          -\tan \\
                                                          \cot
                                                        \end{array}
    \right\}\left(n_0d\right)~.
\end{equation}
Here $m_{\rm e}$ ($n_0$) plays the role of the transverse wave
   number outside (inside) the slab, defined as
\begin{equation}\label{n0me}
    m_{\rm e}^2=\displaystyle\frac{v_{\rm Ae}^2-v^2_{\rm ph}}{v_{\rm
    Ae}^2}k^2~,~~~~~~n^2_0=\displaystyle\frac{\left(v_{\rm ph}-U_0\right)^2
    -v_{\rm A0}^2}    {v_{\rm
    A0}^2    }k^2~.
\end{equation}
In addition, the upper/lower case in Equation~(\ref{flow-dis}) is for
kink/sausage modes.

Finding the period ratios of standing modes supported by a flowing slab is not as straightforward as
   in the static case.
Nevertheless, as detailed in Article~I, a simple graphical means can be used for this purpose by
   capitalizing on the $\omega$--$k$ diagrams.
The component propagating waves in a pair to form standing modes correspond to two curves in an $\omega$--$k$
   diagram, a horizontal cut with a constant $\omega$ would intersect the two at two points.
If the separation between the two points is $2\pi/L$ , then one finds the fundamental mode.
If it is $2n\pi/L$, one finds the $(n-1)^{\rm th}$ overtone.
Let the angular frequency of the fundamental mode be denoted by $\omega_1$,
   that of the $(n-1)^{\rm th}$overtone by $\omega_n$,
   the period ratio is then simply $P_1 /n P_n = \omega_n /n \omega_1$.
In the flowing case, in addition to the density contrast
   $[\rho_0/\rho_{\rm e}]$ and aspect ratio $[d/L]$ of the considered slab,
   the period ratios $[P_1/nP_n]$ depend also on the internal Alfv\'{e}n Mach
   number $[M_A = U_0/v_{A0}]$.

\subsection{Standing Kink Modes}
\begin{figure}
\centering
\includegraphics[totalheight=5.5cm]{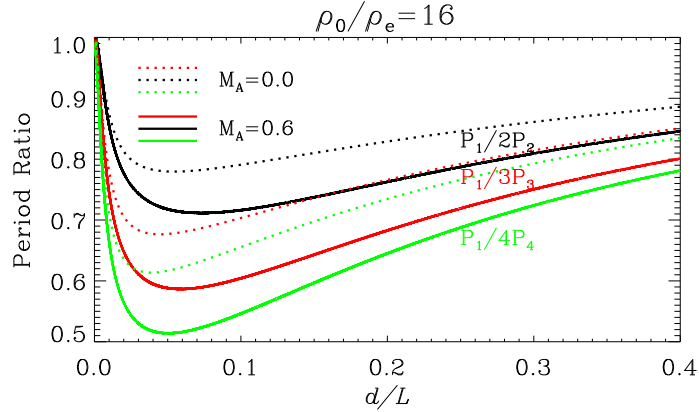}
 \caption{Effects of flow on period ratios $[P_1/nP_n]$ for standing kink modes
    at a density ratio $\rho_0/\rho_{\rm e} = 16$.
Here the period ratios $[P_1/nP_n]$ as functions of aspect ratio $[d/L]$
   are given for a flowing cold slab with an internal Alfv\'en Mach number $M_A=0.6$ (the solid curves),
   and also for a static slab (the dotted curves) for comparison.
The black, red, and green curves describe the period ratios
    $P_1/2P_2$, $P_1/3P_3$, and $P_1/4P_4$.
}
 \label{fig_flow_knk_pr_aspec}
\end{figure}

Figure~\ref{fig_flow_knk_pr_aspec} presents the dependence on the slab aspect ratio $[d/L]$
    of the period ratios $[P_1/nP_n]$ for standing kink modes.
An Alfv\'en Mach number $M_A=0.6$ is chosen (the solid lines) for the magnitude of the internal flow.
The static results are also shown here by the dotted lines for comparison.
One may see that while the trend for the $P_1/nP_n$ profiles is the same in the flowing case
    as in the static one, the effect of the flow is substantial.
Take the minimum value that $P_1/nP_n$ may attain for instance.
For $n=[2,~3,~4]$, this minimum reads $[0.78,~0.676,~0.614]$ in
the static case, but reads $[0.711,~0.587,~0.514]$
    in the case where $M_A=0.6$, amounting to a fractional decrease of {$[8.8\,\%,~13.2\,\%,~16.3\,\%]$}.
{Furthermore, the aspect ratios where the minima are attained tend to decrease with $n$,
    but at a given $n$ they tend to increase by a substantial amount in the flowing case relative to the static one.
When $M_A=0$, they read $[0.053, 0.045, 0.039]$ for $n=[2, 3, 4]$.
The corresponding values are $[0.073, 0.059, 0.05]$ when $M_A=0.6$.
}

\begin{figure}
\centering
 \includegraphics[totalheight=5.5cm]{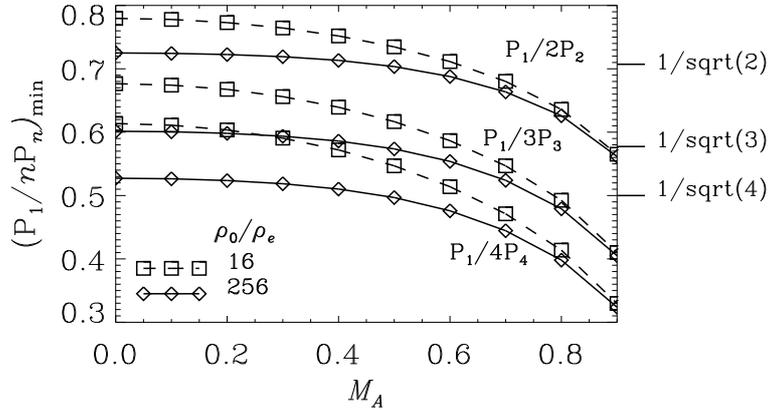}
 \caption{Minima of period ratios $[(P_1/nP_n)_{\rm min}~(n=2,3,4)]$ as functions of
   internal Alfv\'{e}n Mach numbers $[M_A]$.
Two density contrasts $[\rho_0/\rho_{\rm e}]$, 16 and 256,
   are represented by the dashed and solid curves, respectively.
Besides, the symbols (open boxes and diamonds) represent specific computations.
The horizontal bars on the right correspond to $1/\sqrt{n}$,
   the lower bound analytically expected for static slabs.
   }
 \label{fig_flow_kink_pr_min_as_ma}
\end{figure}

The effect of the flow magnitude on the period ratios for standing kink modes
    is better illustrated by Figure~\ref{fig_flow_kink_pr_min_as_ma},
    which presents the dependence on the Alfv\'en Mach number $[M_A]$
    of the minimal period ratios $[(P_1/nP_n)_{\rm min}]$.
Two density contrasts, $\rho_0/\rho_{\rm e} = 16$ and $256$, are examined and plotted by the dashed
   and solid lines, respectively.
The horizontal bars on the right of the panel represent the lower bound $[\sqrt{1/n}]$
   for static slabs.
One finds that the introduction of flow reduces $(P_1/nP_n)_{\rm min}$ as a whole.
When the density contrast $[\rho_0/\rho_{\rm e}]$ is $256$
   for $n$ being $2~(3,~4)$ the minimum of $P_1/nP_n$ reduces from
   $0.73~(0.6,~0.53)$ in the static case to $0.63~(0.48,~0.4)$ at $M_A=0.8$,
   amounting to a fractional reduction of $13.7\,\%~(20.3\,\%,~24.5\,\%)$.
At a lower density contrast $\rho_0/\rho_{\rm e} = 16$,
   the flow effect is even stronger, as demonstrated by that relative to the static case,
   the fractional reduction
   in the minimum of $P_1/2P_2$, $P_1/3P_3$, and $P_1/4P_4$ at $M_A=0.8$
   reads $18.5\,\%$, $27.2\,\%$ and $32.6\,\%$, respectively.
Besides, while for static slabs $P_1/nP_n$
   is subject to the lower limit $1/\sqrt{n}$,
   this is no longer the case for flowing slabs.

The feature that persists in Figure~\ref{fig_flow_kink_pr_min_as_ma}, regardless of $M_A$
   and $\rho_0/\rho_{\rm e}$, is that $P_1/nP_n$ tends to decrease with increasing $n$.
The values $P_1/2P_2 = 0.99$
    and $P_1/3P_3 = 0.965$ measured by~\citet{2009A&A...508.1485V} for a TRACE 171\AA\ loop on 13 May 2001
    agree with this tendency.
As for the extremely small deviation of $P_1/nP_n$ from unity, it may come from the small aspect ratio
    of EUV loops, or may be due to the longitudinal structuring
    in loop density as well as magnetic field strength as suggested by~\citet{2009A&A...508.1485V}.
On the other hand, the NoRH loop that underwent standing kink oscillations as measured by~\citet{2013SoPh..284..559K}
    yields values for $P_1/2P_2$ to be $0.83$ and for $P_1/3P_3$ to be $0.91$, at variance with
    the afore-mentioned tendency.
However, a closer inspection of Figure~5 therein shows that
    the period $11.5$~seconds regarded as the second overtone
    may correspond in fact to even higher overtones, for the spatial distribution
    of the associated spectral power does not clearly show a signature with
    three peaks between the two footpoints.
In contrast, the periods $31.2$~seconds and $21.3$~seconds correspond to distributions of spectral power
    concentrated near the loop apex and close to footpoints, respectively,
    indeed in line with the expectation that these are for the fundamental and the 1st overtone.
If one tentatively attributes $11.5$~seconds to the third overtone, then one finds that $P_1/4P_4 = 0.68$,
    which no longer contradicts the tendency for $P_1/nP_n$ to decrease with $n$.
Given the uncertainties in interpreting this $11.5$-second period,
    the final answer certainly awaits a dedicated calculation, which should
    take into account both longitudinal and transverse structuring.
We note that the latter is necessary given that the NoRH loop in question has an aspect ratio of $0.19$
    for which the dispersion due to finite aspect ratios cannot be neglected.

\begin{figure}
\centering
 \includegraphics[totalheight=5.5cm]{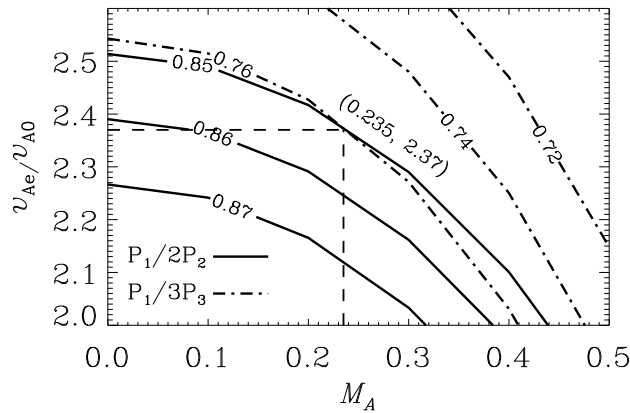}
 \caption{Contours of period ratios [$P_1/2P_2$ (solid) and
   $P_1/3P_3$ (dash--dotted)] with respect to $M_A$ and $v_{\rm Ae}/v_{\rm A0}$
   at an aspect ratio $d/L=0.05$.
The contours for $P_1/2P_2$ ($P_1/3P_3$) are equally spaced
   by $0.01$ ($0.02$). }
 \label{fig_flow_knk_pr_contour}
\end{figure}

That a field-aligned flow may have substantial effects on the period ratios for standing kink modes
    leads naturally to a diagnostic tool for deducing the flow magnitude.
This is illustrated in Figure~\ref{fig_flow_knk_pr_contour}, where the contours of
    period ratios $P_1/2P_2$ (the solid curves) and $P_1/3P_3$ (dash--dotted)
     are shown as a function of the Alfv\'en Mach number
    $[M_{\rm A}]$ and the Alfv\'en speed ratio
    $[v_{\rm Ae}/v_{\rm A0}]$, at a given aspect ratio $[d/L]$.
For illustrative purposes, the results shown are for $d/L$ being $0.05$, which is not unrealistic
    but lies within the range deduced for
    TRACE 171\AA\ loops~\citep[][Table 1]{2002ApJ...576L.153O}.
One sees that at a given density ratio, both $P_1/2P_2$ and $P_1/3P_3$ decrease with increasing $M_A$.
Likewise, at a given $M_A$, both $P_1/2P_2$ and $P_1/3P_3$ decrease with $v_{\rm Ae}/v_{\rm A0}$.
Although following the same pattern, the two sets of contours may intersect at a series of points:
    the solid contour corresponding to $P_1/2P_2 = 0.85$ intersects
    the dash-dotted one corresponding to $P_1/3P_3 = 0.76$ at $(M_A, v_{\rm Ae}/v_{\rm A0}) = (0.235, 2.37)$.
So the point that we make here is that if a loop undergoes standing kink oscillations, and if the oscillating signals
    contain periods of both the fundamental and its first and second overtones,
    then with the measured periods one can readily derive simultaneously the
    density contrast of the loop with its surroundings
    and the internal Alfv\'en Mach number:
    Neither of these two is easy to measure directly from an observational standpoint.
Actually, if from observations one sees also the third overtone, then the measured $P_1/4P_4$
    can provide an additional means for checking the quality of the derived flow magnitude
    and density contrast.
For instance, at this afore-mentioned combination of $[M_A, v_{\rm Ae}/v_{\rm A0}]$, the theoretically expected
    $P_1/4P_4$ at the given $d/L$ {would be $\approx 0.7$}, meaning that
    the amount by which the measured $P_1/4P_4$ deviates from this theoretical value
    serves as a natural uncertainty measure.

\subsection{Standing Sausage Modes}
\begin{figure}
\centering
 \includegraphics[totalheight=5.5cm]{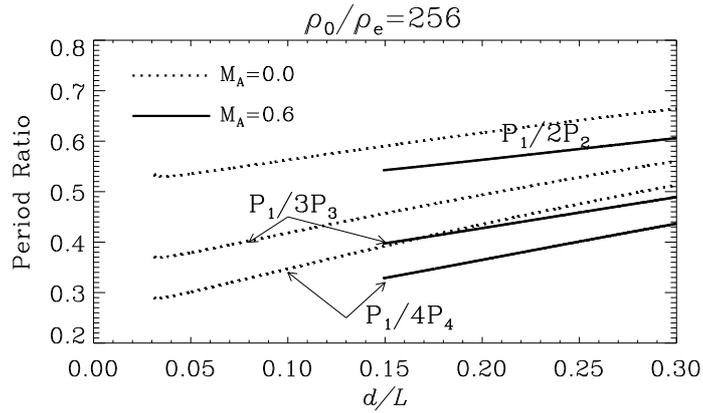}
 \caption{Period ratios $[P_1/nP_n~(n=2,~3,~4)]$ as functions of aspect ratio
    $[d/L]$ for a static (dotted lines)
    and a flowing slab (solid).
}
 \label{fig_flow_saus_pr}
\end{figure}

Now we examine the standing sausage modes supported by a flowing slab.
Figure~\ref{fig_flow_saus_pr} presents $P_1/nP_n$ with $n$ ranging from $2$ to $4$
    as a function of aspect ratio $[d/L]$ at a density contrast of
    $\rho_0/\rho_{\rm e}=256$.
To bring out the flow effect, the static case (dotted lines) is also presented in addition
    to a flowing case where the Alfv\'en Mach number $[M_A]$ is $0.6$.
One can see that all period ratios $[P_1/nP_n]$ decrease with the introduction of the flow.
For instance, at $d/L=0.2$ the period ratios with $n$ being $[2, 3, 4]$ reduce
    from $[0.62, 0.5, 0.44]$ in the static case to
    [0.567, 0.434, 0.372] in the flowing case.
The fractional reduction is therefore $[8.7\,\%, 13.4\,\%, 16.2\,\%]$.
One can see that for a density contrast as large as $256$,
    $P_1/nP_n$ close to the cutoff aspect ratio in the static case is close to
    their lower limits $[1/n]$ established by Equation~(\ref{infty-sausage}).
However, while at a given aspect ratio the flow reduces $P_1/nP_n$ to a substantial extent,
    the minimal $P_1/nP_n$ in the flowing case is similar to that in the static case,
    the reason being that the cutoff
    aspect ratio $[(d/L)_{\rm cutoff}]$ shifts towards larger values  when a flow is present.
One finds that in the static case $(d/L)_{\rm cutoff}$ is
    {$0.031$}.
However it is {$0.148$} in the case where
    $M_A = 0.6$, amounting to an increase by a factor of {$3.8$}.

\begin{figure}
\centering
 \includegraphics[totalheight=5.5cm]{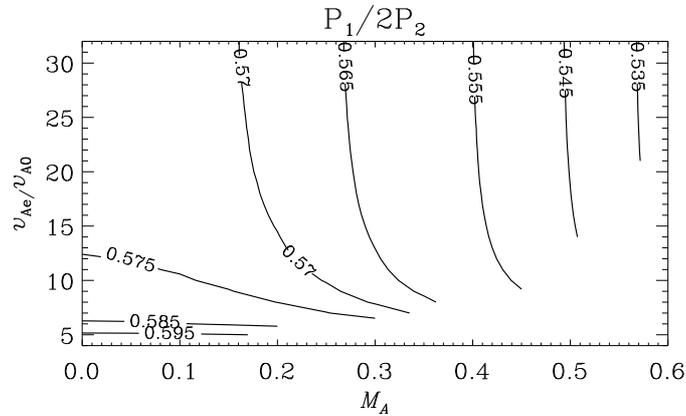}
 \caption{Period ratio $[P_1/2P_2]$ of standing sausage modes for flowing slabs with an aspect ratio of $0.12$
 as a function of the Alfv\'en speed ratio $[v_{Ae}/v_{A0}]$
 and internal Alfv\'en Mach number $[M_A]$.
 This aspect ratio corresponds to the flaring loop that underwent
     a standing sausage oscillation on 12 January 2000 as measured with NoRH~\citep{2003A&A...412L...7N}.
}
 \label{fig_flow_saus_P1P2_given_dL}
\end{figure}

That $(P_1/nP_n)_{\rm min}$ changes little even in the presence of a substantial flow
    raises the question of how to explain the observed period ratios
    as reported by~\citet{2003A&A...412L...7N} for NoRH flaring loops
    and by~\citet{2008MNRAS.388.1899S} for cool H$\alpha$ loops.
Recall that in the former $P_1/2P_2$ reads $0.82$ at an aspect ratio $d/L=0.12$.
Figure~\ref{fig_flow_saus_P1P2_given_dL} examines what we derive for $P_1/2P_2$ at
    this aspect ratio for an extensive range of $v_{Ae}/v_{A0}$ and $M_A$.
The contours of $P_1/2P_2$ are equally spaced by $0.01$.
One can see that the lower-right portion is blank, for at those given $v_{Ae}/v_{A0}$
    trapped sausage modes are not allowed at this aspect ratio.
Looking at the values of $P_1/2P_2$, one sees that $P_1/2P_2$ tends to decrease with $M_A$
    at a given $v_{Ae}/v_{A0}$, and this tendency is stronger for larger $v_{Ae}/v_{A0}$.
Nevertheless, in the whole parameter range $P_1/2P_2$ varies by no more than $\approx 12\,\%$
    if one compares their values at the lower-left corner with those at the upper right one.
More importantly, they never exceed $\approx 0.6$, which is considerably smaller than the observed value.
Actually, given that the theoretically expected values are smaller than the observed,
    introducing a flow makes the comparison even more undesirable
    since a flow would decrease rather than increase $P_1/2P_2$.
A similar study at $d/L=0.03$, pertinent to the H$\alpha$ loops reported by~\citet{2008MNRAS.388.1899S},
    shows that $P_1/2P_2$ does not exceed {$0.53$}, which is far from the measured value
    which is $0.84$.
We conclude that the multiple periods of standing sausage modes measured so far
    remain to be explained.

\begin{figure}
\centering
 \includegraphics[totalheight=5.5cm]{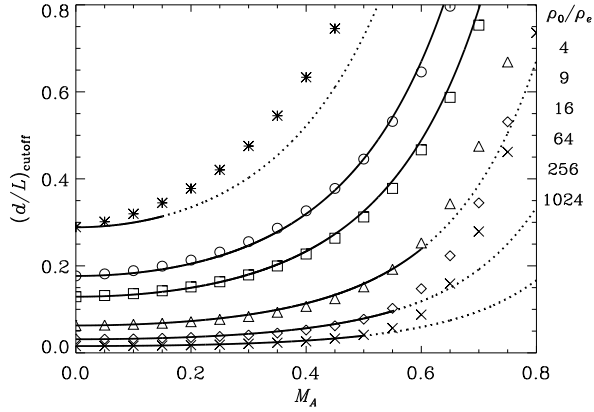}
 \caption{Aspect ratio cutoffs as functions of internal Alfv\'{e}n Mach number $[M_A]$ for a series of
    density contrasts $[\rho_0/\rho_{\rm e}]$.
In addition to the numerically derived values given by the symbols,
    an analytical fit $\left[1/\sqrt{4(\rho_0/\rho_{\rm e}-1)}\right]\exp\left(3.7 M_A^2\right)$
    is given by the curves.
The curves are solid where this fit has an accuracy better than $10\,\%$,
    and are dotted otherwise. }
 \label{fig_flow_saus_cutoff}
\end{figure}

Nevertheless, the sensitive dependence on the Alfv\'en Mach number of the cutoff aspect ratio $[(d/L)_{\rm cutoff}]$
    has a number of observational implications.
To show this, let us first examine this dependence further by conducting a parameter study
    for an extensive range of density contrast $[\rho_0/\rho_{\rm e}]$,
    the result of which is shown in Figure~\ref{fig_flow_saus_cutoff}.
Here the symbols represent $(d/L)_{\rm cutoff}$ as a function of $M_A$ at a series of $\rho_0/\rho_{\rm e}$
    ranging from $4$ to $1024$.
The curves show an analytical fit
\begin{eqnarray}
    (d/L)_{\rm cutoff, fit} =
      \frac{1}{2}\sqrt{\frac{1}{\rho_0/\rho_{\rm e}-1}} \exp\left(3.7 M_A^2\right) .
\label{eq_flow_saus_cut_fit}
\end{eqnarray}
The curves are solid where this fit is better than $10\,\%$ in accuracy and dotted otherwise.
One sees that Equation~(\ref{eq_flow_saus_cut_fit}) provides a good fit to the numerically
    derived $(d/L)_{\rm cutoff}$ for $\rho_0/\rho_{\rm e}$
    ranging from $9$ to $1024$ and $M_A$ in the range of $[0, 0.5]$.

The simplicity of Equation~(\ref{eq_flow_saus_cut_fit}) makes it a convenient means to deduce
    the combination of $(\rho_0/\rho_{\rm e}, M_A)$ pertinent to a loop
    provided that it undergoes a standing sausage oscillation
    and that its aspect ratio is known.
This is because, now that it does show oscillations in this particular mode,
    its aspect ratio has to be larger than determined by Equation~(\ref{eq_flow_saus_cut_fit}).
If the density contrast in this combination is further measured,
    then one readily derives an upper limit of $M_A$.
Assuming $\rho_0/\rho_{\rm e} = 100$, if the loop aspect ratio is $0.12$~\citep{2003A&A...412L...7N},
    then one finds from Equation~(\ref{eq_flow_saus_cut_fit}) that $M_A$ has to be smaller than $0.49$.
If, on the other hand, taking $d/L$ to be $0.03$~\citep{2008MNRAS.388.1899S},
    one finds that if $\rho_0/\rho_{\rm e} = 600$,
    then $M_A$ has to be smaller than $0.32$.
Given that measuring the flow magnitude in loops is a nontrivial task~\citep{2010LRSP....7....5R},
    this simple formula offers a tool for constraining
    the flow magnitude.

\begin{figure}
\centering
 \includegraphics[totalheight=5.5cm]{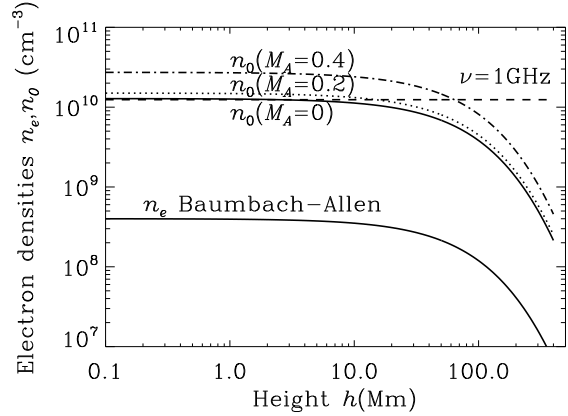}
 \caption{Height dependence of electron densities of loops $[n_0]$ and their surroundings $[n_{\rm e}]$.
 Here $n_{\rm e}$ is given by the Baumbach--Allen empirical model.
 Also given are a number of $n_0$, the minimum loop density for a loop with width-to-length
    ratio of $1/4$ to support sausage modes.
 In addition to the static case, two flowing cases with internal Alfv\'en Mach numbers
    $[M_A]$ being $0.2$ (the dotted curve) and $0.4$ (dash--dotted) are also given.
 For comparison, the density corresponding to a plasma frequency of 1~GHz is plotted by
    the dashed line.}
 \label{fig_partial_saus}
\end{figure}

Actually, Equation~(\ref{eq_flow_saus_cut_fit}) also enables one to take a further look
    at the standing sausage oscillations observed prior to the 2000s
    using radio bands with observing frequencies $\nu \lesssim 1$~GHz.
This was done in~\citet{2004ApJ...600..458A} (hereafter ANM04) who capitalized on the cutoff aspect ratios
    for static loops.
At a given width-to-length ratio, taken to be $1/4$ therein,
    for loops to support standing modes the density ratio has to exceed some critical value.
However, the electron density of the loops $n_0$ as well as that of their
    surroundings $n_{\rm e}$
    are not arbitrary but observationally constrained.
A natural constraint on $n_0$ comes from the observing frequency $[\nu]$ since the emission
    in this range comes primarily at the plasma frequency.
As for $n_{\rm e}$, a wealth of empirical data exists on
    which empirical formulae, such as the Baumbach--Allen one~\citep{2000Cox}, were built.
The point made by ANM04 is that, as $n_{\rm e}$ shows a height dependence, the ratio $n_0/n_{\rm e}$
    (equivalent to $\rho_0/\rho_{\rm e}$) may not exceed the critical value
    throughout the loop but only for a segment of it.
While ANM04 adopt a formula valid for static loops, let us extend the idea
    therein to incorporate the flow effect as well.
Figure~\ref{fig_partial_saus} presents, similar to Figure 4 in ANM04,
    as functions of height $h$,
    the electron densities of the ambient corona $[n_{\rm e}]$
    as well as the minimum loop density $[n_0]$ required for loops with
    a width-to-length ratio $[w/L]$ of $1/4$ to support sausage modes.
Here the Baumbach--Allen density is assumed for $n_{\rm e}$,
    reading $n_{\rm e}(h) \approx 4\times 10^8/(1+h/{\rm R}_\odot)^9$~cm$^{-3}$ with ${\rm R}_\odot$ being
    the solar radius.
On the other hand, $n_0$ is calculated for both the static case $M_A=0$
    and two flowing cases where $M_A=0.2$ (the dotted curve)
    and $0.4$ (dash--dotted), using the formula $n_0/n_{\rm e} = 1+\exp(7.4 M_A^2)/(w/L)^2$.
The latter formula simply follows from Equation~(\ref{eq_flow_saus_cut_fit}).
For comparison, the horizontal dashed line presents the density that corresponds to a
    plasma frequency of $\nu = 1$~GHz, calculated with the expression $(\nu/8980)^2$,
    in which $\nu$ is in Hz and the resulting density is in cm$^{-3}$.
If an $n_0$ curve is below the dashed line, then the sausage mode is termed a ``free'' one
    by ANM04 in that the whole loop may experience the standing oscillations.
However, if only part of an $n_0$ curve is below the dashed line, then only the loop segment that
    has heights above a certain value can experience sausage oscillations, which are termed ``partial'' modes
    by ANM04.
As can be seen from Figure~\ref{fig_partial_saus},
    while for $M_0=0$ this critical height reads $2.5$~Mm,
    it increases substantially to $15$~Mm when the loop flow corresponds to an $M_A$ of $0.2$,
    and further increases to $63.5$~Mm when $M_A=0.4$.
From this we conclude that the flow effect should be taken into account
    when one tries to deduce the extent of the loop segment
    that can support a partial sausage mode.

\section{Summary}
\label{sec_conc}

The present study is motivated by the apparent lack of a detailed investigation into
    the effects of a significant field-aligned flow in coronal loops
    on standing modes that they support in general,
    the period ratios and cutoff aspect ratios in particular.
By period ratios, we mean $P_1/nP_n$ where $P_1$ stands for the period of the fundamental mode,
    and $P_n$ represents the period of its $(n-1)^{\rm th}$ overtone.
The aspect ratio is defined as the ratio of loop half-width $[d]$
    to its length $[L]$.
The Alfv\'en Mach number $[M_A]$, which measures the loop flow speed
    in units of the internal Alfv\'en speed, is also relevant.
Appropriate for coronal environments, the loops are modeled as a zero-$\beta$ magnetic slab, where
    $\beta$ is the ratio of the thermal to magnetic pressure.
Our main results can be summarized as follows.

\begin{enumerate}
 \item
   We presented a detailed analytical analysis of static slabs with transverse density structuring described
      by an Epstein profile.
   The results concern the behavior of $P_1/nP_n$ in general,
      their behavior in the thin- and wide-slab limits in particular.
   By doing so, we generalize the study by~\citet{2011A&A...526A..75M} to overtones of arbitrary order,
      and establish the lower bound that $P_1/nP_n$ may attain.
   Solving the static-slab problem where the density profile is of a step-function form instead,
      we find that these lower limits also hold.
\item
   For standing kink modes supported by flowing slabs, the flow is found to significantly reduce
      the period ratios $[P_1/nP_n]$ for all of the considered density contrasts $[\rho_0/\rho_{\rm e}]$.
   This is true even when $\rho_0/\rho_{\rm e}$ is very large~(the solid curves in Figure~\ref{fig_flow_kink_pr_min_as_ma}),
      in which case while for static slabs $P_1/nP_n$ almost reaches the analytically expected lower limit $[1/\sqrt{n}]$,
      they may be reduced by $[13.7\,\%, 20.3\,\%, 24.5\,\%]$ for $n$ being $[2, 3, 4]$
      for the Alfv\'en Mach number $[M_A]$ in the range of $[0, 0.8]$.
   For lower, and therefore more realistic, density contrasts, the flow effect is even stronger.
\item
   A way of deducing simultaneously $\rho_0/\rho_{\rm e}$ and $M_A$, pertinent to standing kink modes,
       is illustrated in Figure~\ref{fig_flow_knk_pr_contour}.
   The idea is simply that, if the main contributor to the departure of $P_1/nP_n$ from unity is
       the wave dispersion due to transverse structuring in density and flow speeds,
       then at a given aspect ratio, if $P_1/nP_n$ with two different $n$ are measured,
       the combination of $(\rho_0/\rho_{\rm e}, M_A)$ may be readily read out from
       a contour plot similar to Figure~\ref{fig_flow_knk_pr_contour}.
   In fact, if an additional overtone is also measured, an uncertainty measure of the deduced
       $(\rho_0/\rho_{\rm e}, M_A)$ can be readily deduced.
\item
   For standing sausage modes supported by flowing slabs, the flow effect on $P_1/nP_n$ is not as strong but still
       substantial in reducing their values at a given slab aspect ratio.
   However, in the flowing case $P_1/nP_n$ is still bounded by the lower limit $1/n$ established for
       static slabs, the reason being that the flow significantly enhances the cutoff aspect ratio, below which
       sausage modes are no longer trapped.
\item
   A parameter study on the cutoff aspect ratio for standing sausage modes $[(d/L)_{\rm cutoff}]$
       yields that it may be satisfactorily approximated by Equation~(\ref{eq_flow_saus_cut_fit}),
       which involves $\rho_0/\rho_{\rm e}$ and $M_A$ in a simple manner,
       when $\rho_0/\rho_{\rm e}$ is in the range of $[9, 1024]$ and $M_A$ in the range $[0, 0.5]$.
   This simple formula allows one to analytically constrain $\rho_0/\rho_{\rm e}$ and $M_A$ making use of
       only the fact that a loop experiences standing sausage oscillations.
   Provided that the aspect ratio is known, if the density contrast is found as well, then
       one readily derives the upper limit for $M_A$.
\item
   The simple formula of Equation~(\ref{eq_flow_saus_cut_fit}) enables us to deduce that,
       if the Baumbach--Allen model can describe the density of the ambient corona,
       and if the loops in radio observations with observing frequencies lower than $1$~GHz
       as compiled by~\citet{2004ApJ...600..458A} correspond to a width-to-length ratio of $1/4$,
       then only the segment at heights above $15$ ($64$)~Mm can experience a partial sausage oscillation
       when $M_A$ is $0.2$ (0.4), which are substantially higher than
       $2.5$~Mm found in the static case.
       This expands the original discussion on sausage modes confined
        to only a loop segment rather than
       perturbing the entire loop~\citep{2004ApJ...600..458A}.
\end{enumerate}

\begin{acks}
This research is supported by the 973 program 2012CB825601, the National Natural Science Foundation of China (40904047,
41174154, and 41274176), the Ministry of Education of
China (20110131110058 and NCET-11-0305), and by the Provincial Natural Science Foundation
of Shandong via Grant JQ201212.
\end{acks}


\end{article}
\end{document}